\pdfoutput=1

\documentclass{llncs}
\usepackage{graphicx}

\usepackage{times}
\usepackage{mathptm}
\usepackage{cite}
\usepackage{amsmath}

\DeclareSymbolFont{AMSb}{U}{msb}{m}{n}
\DeclareSymbolFontAlphabet{\Bbb}{AMSb}

\makeatletter
\def\hb@xt@{\hbox to }
\makeatother

\let\oldendproof\endproof
\def\endproof{\qed\oldendproof}

\begin{document}
\title{The $h$-Index of a Graph and its\\ Application to Dynamic Subgraph Statistics} 

\author{David Eppstein\inst{1} and Emma S. Spiro\inst{2}}

\institute{Computer Science Department, University of California, Irvine\and
Department of Sociology, University of California, Irvine}

\maketitle   

\begin{abstract}
We describe a data structure that maintains the number of triangles in a dynamic undirected graph, subject to insertions and deletions of edges and of degree-zero vertices. More generally it can be used to maintain the number of copies of each possible three-vertex subgraph in time $O(h)$ per update, where $h$ is the \emph{$h$-index} of the graph, the maximum number such that the graph contains $h$ vertices of degree at least $h$. We also show how to maintain the $h$-index itself, and a collection of $h$ high-degree vertices in the graph, in constant time per update. Our data structure has applications in social network analysis using the exponential random graph model (ERGM); its bound of $O(h)$ time per edge is never worse than the $\Theta(\sqrt m)$ time per edge necessary to list all triangles in a static graph, and is strictly better for graphs obeying a power law degree distribution. In order to better understand the behavior of the $h$-index statistic and its implications for the performance of our algorithms, we also study the behavior of the $h$-index on a set of 136 real-world networks.
\end{abstract}

\section{Introduction}
\label{sec:intro}

The \emph{exponential random graph model} (ERGM, or $p^*$ model) \cite{Fra-SN-91,WasPat-Psy-96,RobMor-SN-07} is a general technique for assigning probabilities to graphs that can be used both to generate simulated data for social network analysis and to perform probabilistic reasoning on real-world data. In this model, one fixes the vertex set of a graph, identifies certain \emph{features} $f_i$ in graphs on that vertex set, determines a \emph{weight} $w_i$ for each feature, and sets the probability of each graph $G$  to be proportional to an exponential function of the sum of its features' weights, divided by a normalizing constant $Z$:
$${\rm Pr}(G)=\frac{\exp\sum_{f_i\in G} w_i}{Z}.$$
$Z$ is found by summing over all graphs on that vertex set:
$$Z=\sum_G \exp\sum_{f_i\in G} w_i.$$
For instance, if each potential edge is considered to be a feature and all edges have weight $\ln\frac{p}{1-p}$, the normalizing constant $Z$ will be $(1-p)^{-n(n-1)/2}$, and the
probability of any particular $m$-edge graph will be $p^m(1-p)^{n(n-1)/2-m}$, giving rise to the familiar Erd\H{o}s-R\'enyi $G(n,p)$ model. However, the ERG model is much more general than the Erd\H{o}s-R\'enyi model: for instance, an ERGM in which the features are whole graphs can represent arbitrary probabilities. The generality of this model, and its ability to define probability spaces lacking the independence properties of the simpler Erd\H{o}s-R\'enyi model, make it difficult to analyze analytically. Instead, in order to generate graphs in an ERG model or to perform other forms of probabilistic reasoning with the model, one typically uses a Markov Chain Monte Carlo method~\cite{Sni-JoSS-02} in which one performs a large sequence of small changes to sample graphs, updates after each change the counts of the number of features of each type and the sum of the weights of each feature, and uses the updated values to determine whether to accept or reject each change. Because this method must evaluate large numbers of graphs, it is important to develop very efficient algorithms for identifying the features that are present in each graph.

Typical features used in these models take the form of small subgraphs:  \emph{stars} of several edges with a common vertex (used to represent constraints on the degree distribution of the resulting graphs), \emph{triangles} (used in the triad model~\cite{FraStr-JASA-86}, an important predecessor of ERG models, to represent the likelihood that friends-of-friends are friends of each other), and more complicated subgraphs used to control the tendencies of simpler models to generate unrealistically extremal graphs~\cite{SniPatRob-SM-06}. Using highly local features of this type is important for reasons of computational efficiency, matches well the type of data that can be obtained for real-world social networks, and is well motivated by the local processes believed to underly many types of social network. Thus, ERGM simulation leads naturally to problems of \emph{subgraph isomorphism}, listing or counting all copies of a given small subgraph in a larger graph.

There has been much past algorithmic work on subgraph isomorphism problems. It is known, for instance, that an $n$-vertex graph with $m$ edges may have $\Theta(m^{3/2})$ triangles and four-cycles, and all triangles and four-cycles can be found in time $O(m^{3/2})$~\cite{ItaRod-SJC-78,ChiNis-SICOMP-85}. All cycles of length up to seven can be counted rather than listed in time of $O(n^\omega)$~\cite{AloYusZwi-Algo-97} where $\omega\approx 2.376$ is the exponent from the asymptotically fastest known matrix multiplication algorithms~\cite{CopWin-JSC-90}; this improves on the previous $O(m^{3/2})$ bounds for dense graphs. Fast matrix multiplication has also been used for more general problems of finding and counting small cliques in graphs and hypergraphs~\cite{EisGra-TCS-04,KloKraMue-IPL-00,NesPol-CMUC-85,VasWil-STOC-09,Yus-IPL-06}.
In planar graphs, or more generally graphs of bounded local treewidth, the number of copies of any fixed subgraph may be found in linear time~\cite{Epp-JGAA-99,Epp-Algo-00},
even though this number may be a large polynomial of the graph size~\cite{Epp-JGT-93}. Approximation algorithms for subgraph isomorphism counting problems based on random sampling have also been studied, with motivating applications in bioinformatics~\cite{DukLefRod-SJC-95,KatItzMil-BI-04,PrzCorJur-BI-06}.
However, much of this subgraph isomorphism research makes overly restrictive assumptions about the graphs that are allowed as input, runs too slowly for the ERGM application, depends on impractically complicated matrix multiplication algorithms, or does not capture the precise subgraph counts needed to accurately perform Markov Chain Monte Carlo simulations.

Markov Chain Monte Carlo methods for ERGM-based reasoning process a sequence of graphs each differing by a small change from a previous graph, so it is natural to seek additional efficiency by applying \emph{dynamic graph algorithms} \cite{EppGalIta-ATCH-99,FeiKan-HDCM-00,ThoKar-SWAT-00}, data structures to efficiently maintain properties of a graph subject to vertex and edge insertions and deletions. However, past research on dynamic graph algorithms has focused on problems of connectivity, planarity, and shortest paths, and not on finding the features needed in ERGM calculations.
In this paper, we apply dynamic graph algorithms to subgraph isomorphism problems important in ERGM feature identification. To our knowledge, this is the first work on dynamic algorithms for subgraph isomorphism.

A key ingredient in our algorithms is the $h$-index, a number introduced by Hirsch~\cite{Hir-PNAS-05} as a way of balancing prolixity and impact in measuring the academic achievements of individual researchers. Although problematic in this application~\cite{AdlEwiTay-JCQAR-08}, the $h$-index can be defined and studied mathematically, in graph-theoretic terms, and provides a convenient measure of the uniformity of distribution of edges in a graph. Specifically, for a researcher, one may define a bipartite graph in which the vertices on one side of the bipartition represent the researcher's papers, the vertices on the other side represent others' papers, and edges correspond to citations by others of the researcher's papers. The $h$-index of the researcher is the maximum number $h$ such that at least $h$ vertices on the researcher's side of the bipartition each have degree at least $h$. We generalize this to arbitrary graphs, and define the $h$-index of any graph to be the maximum $h$ such that the graph contains $h$ vertices of degree at least $h$. Intuitively, an algorithm whose running time is bounded by a function of $h$ is capable of tolerating arbitrarily many low-degree vertices without slowdown, and is only mildly affected by the presence of a small number of very high degree vertices; its running time depends primarily on the numbers of intermediate-degree vertices. As we describe in more detail in Section~\ref{sec:data}, the $h$-index of any graph with $m$ edges and $n$ vertices is sandwiched between $m/n$ and $\sqrt{2m}$, so it is sublinear whenever the graph is not dense, and the worst-case graphs for these bounds have an unusual degree distribution that is unlikely to arise in practice.

Our main result is that we may maintain a dynamic graph, subject to edge insertions, edge deletions, and insertions or deletions of isolated vertices, and maintain the number of triangles in the graph, in time $O(h)$ per update where $h$ is the $h$-index of the graph at the time of the update. This compares favorably with the time bound of $\Theta(m^{3/2})$ necessary to list all triangles in a static graph. In the same $O(h)$ time bound per update we may more generally maintain the numbers of three-vertex induced subgraphs of each possible type, and in constant time per update we may maintain the $h$-index itself.
Our algorithms are randomized, and our analysis of them uses amortized analysis to bound their expected times on worst-case input sequences. Our use of randomization is limited, however, to the use of hash tables to store and retrieve data associated with keys in $O(1)$ expected time per access. By using either direct addressing or deterministic integer searching data structures instead of hash tables we may avoid the use of randomness at an expense of either increased space complexity or an additional factor of $O(\log\log n)$ in time complexity; we omit the details.

We also study the behavior of the $h$-index, both on scale-free graph models and on a set of real-world graphs used in social network analysis. We show that for scale-free graphs, the $h$-index scales as a power of $n$, less than its square root, while in the real-world graphs we studied the scaling exponent appears to have a bimodal distribution.


\section{Dynamic $h$-Indexes of Integer Functions}
\label{sec:dynh}

We begin by describing a data structure for the following problem, which generalizes that of maintaining $h$-indexes of dynamic graphs. We are given a set $S$, and a function $f$ from $S$ to the non-negative integers, both of which may vary discretely through a sequence of updates: we may insert or delete elements of $S$ (with arbitrary function values for the inserted elements), and we may make arbitrary changes to the function value of any element of $S$. As we do so, we wish to maintain a set $H$ such that, for every $x\in H$, $f(x)\ge |H|$, with $H$ as large as possible with this property. We call $|H|$ the \emph{$h$-index} of $S$ and $f$, and we call the partition of $S$ into the two subsets $(H,S\setminus H)$ an \emph{$h$-partition} of $S$ and $f$.

To do so, we maintain the following data structures:
\begin{itemize}
\item A dictionary $F$ mapping each $x\in S$ to its value under $f$: $F[x]=f(x)$.
\item The set $H$ (stored as a dictionary mapping members of $H$ to an arbitrary value).
\item The set $B=\{x\in H\mid f(x)=|H|\}$.
\item A dictionary $C$ mapping each non-negative integer $i$ to the set $\{x\in S\setminus B\mid f(x)=i\}$. We only store these sets when they are non-empty, so the situation that there is no $x$ with $f(x)=i$ can be detected by the absense of $i$ among the keys of $C$.
\end{itemize}

To insert an element $x$ into our structure, we first set $F[x]=f(x)$, and add $x$ to $C[f(x)]$ (or add a new set $\{x\}$ at $C[f(x)]$ if there is no existing entry for $f(x)$ in $C$). Then, we test whether $f(x)> |H|$. If not, the $h$-index does not change, and the insertion operation is complete. But if $f(x)>|H|$, we must include $x$ into $H$. If $B$ is nonempty, we choose an arbitrary $y\in B$, remove $y$ from $B$ and from $H$, and add $y$ to $C[|H|]$ (or create a new set $\{y\}$ if there is no entry for $|H|$ in $C$). Finally, if $f(x)>|H|$ and $B$ is empty, the insertion causes the $h$-index ($|H|$) to increase by one. In this case, we test whether there is an entry for the new value of $|H|$ in $C$. If so, we set $B$ to equal the identity of the set in $C[|H|]$ and delete the entry for $|H|$ in $C$; otherwise, we set $B$ to the empty set.

To remove $x$ from our structure, we remove its entry from $F$ and we remove it from $B$ (if it belongs there) or from the appropriate set in $C[f(x)]$ otherwise. If $x$ did not belong to $H$, the $h$-index does not change, and the deletion operation is complete. Otherwise, let $h$ be the value of $|H|$ before removing $x$. We remove $x$ from $H$, and attempt to restore the lost item from $H$ by moving an element from $C[h]$ to $B$ (deleting $C[h]$ if this operation causes it to become empty). But if $C$ has no entry for $h$, the $h$-index decreases; in this case we store the identity of set $B$ into $C[h]$, and set $B$ to be the empty set.

Changing the value of $f(x)$ may be accomplished by deleting $x$ and then reinserting it, with some care so that we do not update $H$ if $x$ was already in $H$ and both the old and new values of $f(x)$ are at least equal to $|H|$.

\begin{theorem}
\label{thm:hindex}
The data structure described above maintains the $h$-index of $S$ and $f$, and an $h$-partition of $S$ and $f$, in constant time plus a constant number of dictionary operations per update.
\end{theorem}

We defer the proof to an appendix.

\section{Gradual Approximate $h$-Partitions}
\label{sec:gradual}

Although the vector $h$-index data structure of the previous section allows us to maintain the $h$-index of a dynamic graph very efficiently, it has a property that would be undesirable were we to use it directly as part of our later dynamic graph data structures: the $h$-partition $(H,S\setminus H)$ changes too frequently. Changes to the set $H$ will turn out to be such an expensive operation that we only wish them to happen, on average, $O(1/h)$ times per update. In order to achieve such a small amount of change to $H$, we need to restrict the set of updates that are allowed: now, rather than arbitrary changes to $f$, we only allow it to be incremented or decremented by a single unit, and we only allow an element $x$ to be inserted or deleted when $f(x)=0$. We now describe a modification of the $H$-partition data structure that has this property of changing more gradually for this restricted class of updates.

Specifically, along with all of the structures of the $H$-partition, we maintain a set $P\subset H$ describing a partition $(P,S \setminus P)$. When an element of $x$ is removed from $H$, we remove it from $P$ as well, to maintain the invariant that $P\subset H$. However, we only add an element $x$ to $P$ when an update (an increment of $f(x)$ or decrement of $f(y)$ for some other element $y$) causes $f(x)$ to become greater than or equal to $2|H|$. The elements to be added to $P$ on each update may be found by maintaining a dictionary, parallel to $C$, that maps each integer $i$ to the set $\{x\in H\setminus P\mid f(x)=i\}$.

\begin{theorem}
\label{thm:gradual}
Let $\sigma$ denote a sequence of operations to the data structure described above, starting from an empty data structure. Let $h_t$ denote the value of $h$ after $t$ operations, and let $q=\sum_i 1/h_i$. Then the data structure undergoes $O(q)$ additions and removals of an element to or from $P$.
\end{theorem}

We defer the proof to an appendix. For our later application of this technique as a subroutine in our triangle-finding data structure, we will need a more local analysis. We may divide a sequence of updates into \emph{epochs}, as follows: each epoch begins when the $h$-index reaches a value that differs from the value at the beginning of the previous epoch by a factor of two or more. Then, by Lemma~\ref{lem:quadratic-increments}, an epoch with $h$ as its initial $h$-index lasts for at least $\Omega(h^2)$ steps. Due to this length, we may assign a full unit of credit to each member of $P$ at the start of each epoch, without changing the asymptotic behavior of the total number of credits assigned over the course of the algorithm. With this modification, it follows from the same analysis as above that, within an epoch of $s$ steps, with an $h$-index of $h$ at the start of the epoch, there are $O(s/h)$ changes to $P$.

\section{Counting Triangles}
\label{sec:triangles}

We are now ready to describe our data structure for maintaining the number of triangles in a dynamic graph. It consists of the following information:
\begin{itemize}
\item A count of the number of triangles in the current graph
\item A set $E$ of the edges in the graph, indexed by the pair of endpoints of the edge, allowing constant-time tests for whether a given pair of endpoints are linked by an edge.
\item A partition of the graph vertices into two sets $H$ and $V\setminus H$ as maintained by the data structure from Section~\ref{sec:gradual}.
\item A dictionary $P$ mapping each pair of vertices $u,v$ to a number $P[u,v]$, the number of two-edge paths from $u$ to $v$ via a vertex of $V\setminus H$. We only maintain nonzero values for this number in $P$; if there is no entry in $P$ for the pair $u,v$ then there exist no two-edge paths via $V\setminus H$ that connect $u$ to $v$.
\end{itemize}

\begin{theorem}
\label{thm:triangles}
The data structure described above requires space $O(mh)$ and may be maintained in $O(h)$ randomized amortized time per operation, where $h$ is the $h$-index of the graph at the time of the operation.
\end{theorem}

\begin{proof}
Insertion and deletion of vertices with no incident edges requires no change to most of these data structures, so we concentrate our description on the edge insertion and deletion operations.

To update the count of triangles, we need to know the number of triangles $uvw$ involving the edge $uv$ that is being deleted or inserted. Triangles in which the third vertex $w$ belongs to $H$ may be found in time $O(h)$ by testing all members of $H$, using the data structure for $E$ to test in constant time per member whether it forms a triangle. Triangles in which the third vertex $w$ does not belong to $H$ may be counted in time $O(1)$ by a single lookup in $P$.

The data structure for $E$ may be updated in constant time per operation, and the partition into $H$ and $V\setminus H$ may be maintained as described in the previous sections in constant time per operation. 
Thus, it remains to describe how to update $P$. If we are inserting an edge $uv$, and $u$ does not belong to $H$, it has at most $2h$ neighbors; we examine all other neighbors $w$ of $u$ and for each such neighbor increment the counter in $P[v,w]$ (or create a new entry in $P[v,w]$ with a count of $1$ if no such entry already exists). Similarly if $v$ does not belong to $H$ we examine all other neighbors $w$ of $v$ and for each such neighbor increment $P[u,w]$. If we are deleting an edge, we similarly decrement the counters or remove the entry for a counter if decrementing it would leave a zero value. Each update involves incrementing or decrementing $O(h)$ counters and therefore may be implemented in $O(h)$ time.

Finally, a change to the graph may lead to a change in $H$, which must be reflected in $P$. If a vertex $v$ is moved from $H$ to $V\setminus H$, we examine all pairs $u,w$ of neighbors of $v$ and increment the corresponding counts in $P[u,w]$, and if a vertex $v$ is moved from $V\setminus H$ to $H$ we examine all pairs $u,w$ of neighbors of $v$ and decrement the corresponding counts in $P[u,w]$. This step takes time $O(h^2)$, because $v$ has $O(h)$ neighbors when it is moved in either direction, but as per the analysis in Section~\ref{sec:gradual} it is performed an average of $O(1/h)$ times per operation, so the amortized time for updates of this type, per change to the input graph, is $O(h)$.

The space for the data structure is $O(m)$ for $E$, $O(n)$ for the data structure that maintains $H$, and $O(mh)$ for $P$ because each edge of the graph belongs to $O(h)$ two-edge paths through low-degree vertices.
\end{proof}

\section{Subgraph Multiplicity}
\label{sec:subgraphs}

Although the data structure of Theorem~\ref{thm:triangles} only counts the number of triangles in a graph, it is possible to use it to count the number of three-vertex subgraphs of all types, or the number of induced three-vertex subgraphs of all types. In what follows we let $p_i=p_i(G)$ denote the number of paths of length $i$ in $G$, and we let $c_i=c_i(G)$ denote the number of cycles of length $i$ in $G$.

The set of all edges in a graph $G$ among a subset of three vertices $\{u,v,w\}$ determine one of four possible induced subgraphs: an independent set with no edges, a graph with a single edge, a two-star consisting of two edges, or a triangle. Let $g_0$, $g_1$, $g_2$, and $g_3$ denote the numbers of three-vertex subgraphs of each of these types, where $g_i$ counts the three-vertex induced subgraphs that have $i$ edges.

Observe that it is trivial to maintain for a dynamic graph, in constant time per operation, the three quantities $n$, $m$, and $p_2$, where $n$ denotes the number of vertices of the graph, $m$ denotes the number of edges, and $p_2$ denotes the number of two-edge paths that can be formed from the edges of the graph. Each change to the graph increments or decrements $n$ or $m$. Additionally, adding an edge $uv$ to a graph where $u$ and $v$ already have $d_u$ and $d_v$ incident edges respectively increases $p_2$ by $d_u+d_v$, while removing an edge $uv$ decreases $p_2$ by $d_u+d_v-2$. Letting  $c_3$ denote the number of triangles in the graph as maintained by Theorem~\ref{thm:triangles}, the quantities described above satisfy the matrix equation
$$
\left[\begin{array}{cccc}1&1&1&1\\ 0&1&2&3\\ 0&0&1&3\\ 0&0&0&1\end{array}\right]
\left[\begin{array}{cccc}g_0\\g_1\\g_2\\g_3\end{array}\right]=
\left[\begin{array}{cccc}n(n-1)(n-2)/6\\ m(n-2)\\ p_2\\ c_3\end{array}\right].
$$
Each row of the matrix corresponds to a single linear equation in the $g_i$ values.
The equation from the first row, $g_0+g_1+g_2+g_3=\binom{n}{3}$, can be interpreted as stating that all triples of vertices form one graph of one of these types. The equation from the second row, $g_1+2g_2+3g_3=m(n-2)$, is a form of double counting where the number of edges in all three-vertex subgraphs is added up on the left hand side by subgraph type and on the right hand side by counting the number of edges ($m$) and the number of triples each edge participates in ($n-2$). The third row's equation, $g_2+3g_3=p_2$, similarly counts incidences between two-edge paths and triples in two ways, and the fourth equation $g_3=c_3$ follows since each three vertices that are connected in a triangle cannot form any other induced subgraph than a triangle itself.

By inverting the matrix we may reconstruct the $g$ values:
\begin{eqnarray*}
g_3&=&c_3\\
g_2&=&p_2-3g_3\\
g_1&=&m(n-2)-(2g_2+3g_3)\\
g_0&=&\binom{n}{3}-(g_1+g_2+g_3).
\end{eqnarray*}
Thus, we may maintain each number of induced subgraphs $g_i$ in the same asymptotic time per update as we maintain the number of triangles in our dynamic graph. The numbers of subgraphs of different types that are not necessarily induced are even easier to recover: the number of three-vertex subgraphs with $i$ edges is given by the $i$th entry of the vector on the right hand side of the matrix equation.

As we detail in an appendix, it is also possible to maintain efficiently the numbers of star subgraphs of a dynamic graph, and the number of four-vertex paths in a dynamic graph.

\section{Weighted Edges and Colored Vertices}
\label{sec:weights}

It is possible to generalize our triangle counting method to problems of weighted triangle counting: we assign each edge $uv$ of the graph a weight $w_{uv}$, define the weight of a triangle to be the product of the weights of its edges, and maintain the total weight of all triangles. For instance, if $0\le w_{uv}\le 1$ and each edge is present in a subgraph with probability $w_{uv}$, then the total weight gives the expected number of triangles in that subgraph.

\begin{theorem}
The total weight of all triangles in a weighted dynamic graph, as described above, may be maintained in time $O(h)$ per update.
\end{theorem}

\begin{proof}
We modify the structure $P[u,v]$ maintained by our triangle-finding data structure, so that it stores the weight of all two-edge paths from $u$ to $v$. Each update of an edge $uv$ in our structure involves a set of individual triangles $uvx$ involving vertices $x\in H$ (whose weight is easily calculated) together with the triangles formed by paths counted in $P[u,v]$ (whose total weight is $P[u,v]w_{uv}$). The same time analysis from Theorem~\ref{thm:triangles} holds for this modified data structure.
\end{proof}

For social networking ERGM applications, an alternative generalization may be appropriate. Suppose that the vertices of the given dynamic graph are colored; we wish to maintain the number of triangles with each possible combination of colors. For instance, in graphs representing sexual contacts~\cite{LilEdlAma-Nat-01}, edges between individuals of the same sex may be less frequent than edges between individuals of opposite sexes; one may model this in an ERGM by assigning the vertices two different colors according to whether they represent male or female individuals and using feature weights that depend on the colors of the vertices in the features. As we now show, problems of counting colored triangles scale well with the number of different groups into which the vertices of the graph are classified.

\begin{theorem}
Let $G$ be a dynamic graph in which each vertex is assigned one of $k$ different colors. Then we may maintain the numbers of triangles in $G$ with each possible combination of colors, in time $O(h+k)$ per update.
\end{theorem}

\begin{proof}
We modify the structure $P[u,v]$ stored by our triangle-finding data structure, to store a vector of $k$ numbers: the $i$th entry in this vector records the number of two-edge paths from $u$ to $v$ through a low-degree vertex with color $i$. Each update of an edge $uv$ in our structure involves a set of individual triangles $uvx$ involving vertices $x\in H$ (whose colors are easily observed) together with the triangles formed by paths counted in $P[u,v]$ (with $k$ different possible colorings, recorded by the entries in the vector $P[u,v]$). Thus, the part of the update operation in which we compute the numbers of triangles for which the third vertex has low degree, by looking up $u$ and $v$ in $P$, takes time $O(k)$ instead of $O(1)$. The same time analysis from Theorem~\ref{thm:triangles} holds for all other aspects of this modified data structure.
\end{proof}

Both the weighting and coloring generalizations may be combined with each other without loss of efficiency.

\section{How Small is the $h$-Index of Typical Graphs?}
\label{sec:data}

It is straightforward to identify the graphs with extremal values of the $h$-index.
A split graph in which an $h$-vertex clique is augmented by adding $n-h$ vertices, each connected only to the vertices in the clique, has $n$ vertices and $m=h(n-1)$ edges, achieving an $h$-index of $m/(n-1)$. This is the minimum possible among any graph with $n$ vertices and $m$ edges: any other graph may be transformed into a split graph of this type, while increasing its number of edges and not decreasing $h$, by finding an $h$-partition $(H,V\setminus H)$ and repeatedly replacing edges that do not have an endpoint in $H$ by edges that do have such an endpoint. The graph with the largest $h$-index is a clique with $m$ edges together with enough isolated vertices to fill out the total to  $n$; its $h$-index is $\sqrt{2m}(1+o(1))$. Thus, for sparse graphs in which the numbers of edges and vertices are proportional to each other, the $h$-index may be as small as $O(1)$ or as large as $\Omega(\sqrt n)$. At which end of this spectrum can we expect to find the graphs arising in social network analysis?

One answer can be provided by fitting mathematical models of the \emph{degree distribution}, the relation between the number of incident edges at a vertex and the number of vertices with that many edges, to social networks.
For many large real-world graphs, observers have reported \emph{power laws} in which the number of vertices with degree $d$ is proportional to $nd^{-\gamma}$ for some constant $\gamma>1$; a network with this property is called \emph{scale-free}~\cite{AlbJeoBar-Nat-99,LilEdlAma-Nat-01,New-SIAM-03,Pri-Sci-65}. Typically, $\gamma$ lies in or near the interval $2\le\gamma\le 3$ although more extreme values are possible. The $h$-index of these graphs may be found by solving for the $h$ such that $h=nh^{-\gamma}$; that is, $h=\Theta(n^{1/(1+\gamma)})$. For any $\gamma>1$ this is an asymptotic improvement on the worst-case $O(\sqrt n)$ bound for graphs without power-law degree distributions. For instance, for $\gamma=2$ this would give a bound of $h=O(n^{1/3})$ while for $\gamma=3$ it would give $h=O(n^{1/4})$. That is, by depending on the $h$-index as it does, our algorithm is capable of taking advantage of the extra structure inherent in scale-free graphs to run more quickly for them than it does in the general case.

To further explore $h$-index behavior in real-world networks, we computed the $h$-index for a collection of 136 network data sets typical of those used in social network analysis. These data sets were drawn from a variety of sources traditionally viewed as common repositories for such data. The majority of our data sets were from the well known Pajek datasets \cite{pajek}. Pajek is a program used for the analysis and visualization of large networks. The collection of data available with the Pajek software includes citation networks, food-webs, friendship network, etc. In addition to the Pajek data sets, we included network data sets from UCINET \cite{ucinet}. Another software package developed for network analysis, UCINET includes a corpus of data sets that are more traditional in the social sciences. Many of these data sets represent friendship or communication relations; UCINET also includes various social networks for non-human animals. We also used network data included as part of the statnet software suite \cite{statnet}, statistical  modeling software in R. statnet includes ERGM functionality, making it a good example for data used specifically in the context of ERG models.  Finally, we included data available on the UCI Network Data Repository \cite{datalab}, including some larger networks such as the WWW, blog networks, and other online social networks. By using this data we hope to understand how the $h$-index scales in real-world networks.

Details of the statistics for these networks are presented in an appendix;
a summary of the statistics for network size and $h$-index are in Table~\ref{tab:NetworkSummaryStatistics}, below. For this sample of 136 real-world networks, the $h$-index ranges from 2 to 116. The row of summary statistics for $\log h/\log n$ suggests that, for many networks, $h$ scales as a sublinear power of~$n$. The one case with an $h$-index of 116 represents the ties among Slovenian magazines and journals between 1999 and 2000. The vertices of this network represent journals, and undirected edges between journals have an edge weight that represents the number of shared readers of both journals; this network also includes self-loops describing the number of all readers that read this journal.  Thus, this is a dense graph, more appropriately handled using statistics involving the edge weights than with combinatorial techniques involving the existence or nonexistence of triangles. However, this is the only network from our dataset with an $h$-index in the hundreds. Even with significantly larger networks, the $h$-index appears to scale sublinearly in most cases.

\begin{table}[hbt]
	\centering
		\begin{tabular}{|c|c|c|c|c|}
		\hline
		& min. & median & mean & max. \\
		\hline
		\hline
\quad network size ($n$)\quad & 10  &   67 &  535.3 &  10616 \\
$h$-index ($h$) & 2  &  12  & 19.08  &  116 \\
$\log n$&   2.303   &  4.204  & 4.589  &  9.270\\
$\log h$&	0.6931 &  2.4849 & 2.6150 & 4.7536 \\
$\log h/\log n$&\hbox{\quad 0.2014\quad} &\hbox{\quad 0.6166\quad}&\hbox{\quad 0.6006\quad}&\hbox{\quad 1.0000\quad} 	\\
\hline	
		\end{tabular}
         \smallskip
	\caption{Summary statistics for real-world network data}
	\label{tab:NetworkSummaryStatistics}
\end{table}

\begin{figure}[hbt]
	\centering
		\includegraphics[scale=.4]{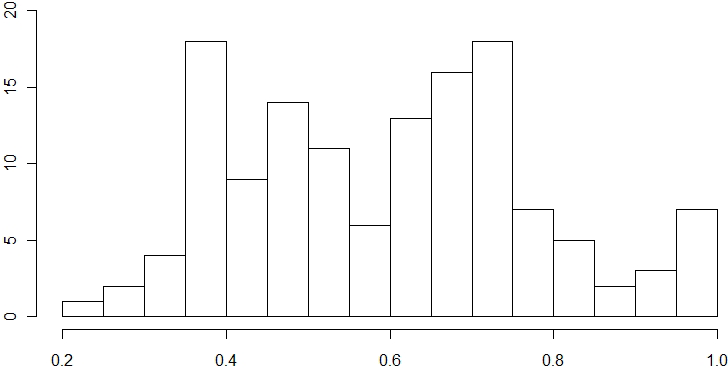}
	\caption{A frequency histogram for $\log h/\log n$.}
	\label{fig:hist}
\end{figure}

A histogram of the $h$-index data in Figure \ref{fig:hist} clearly shows a bimodal distribution. Additionally, as the second peak of the bimodal distribution corresponds to a scaling exponent greater than 0.5, the graphs corresponding to that peak do not match the predictions of the scale-free model. However we were unable to discern a pattern to the types of networks with smaller or larger $h$-indices, and do not speculate on the reasons for this bimodality. We look more deeply at the scaling of the $h$-index using standard regression techniques in an appendix.

\section{Discussion}
\label{sec:discuss}

We have defined an interesting new graph invariant, the $h$-index, presented efficient dynamic graph algorithms for maintaining the $h$-index and, based on them, for maintaining the set of triangles in a graph, and studied the scaling behavior of the $h$-index both on theoretical scale-free graph models and on real-world network data.

There are many directions for future work. For sparse graphs, the $h$-index may be larger than the \emph{arboricity}, a graph invariant used in static subgraph isomorphism~\cite{ChiNis-SICOMP-85,Epp-IPL-94}; can we speed up our dynamic algorithms to run more quickly on graphs of bounded arboricity? We handle undirected graphs but the directed case is also of interest. We would like to find efficient data structures to count larger subgraphs such as 4-cycles, 4-cliques, and claws; dynamic algorithms for these problems are likely to be slower than our triangle-finding algorithms but may still provide speedups over static algorithms. Another network statistic related to triangle counting is the clustering coefficient of a graph; can we maintain it efficiently? Additionally, there is an opportunity for additional work in implementing our data structures and testing their efficiency in practice.

\subsubsection*{Acknowledgements}

This work was supported in part by NSF grant
0830403 and by the Office of Naval Research under grant
N00014-08-1-1015.

{\raggedright
\bibliographystyle{abbrv}
\bibliography{hindex}}

\vfill\eject
\section*{Appendix I: Proof of Theorems \ref{thm:hindex} and \ref{thm:gradual}}

We begin by proving Theorem~\ref{thm:hindex}, the correctness of our data structure for maintaining the $h$-index and $h$-partition, and the analysis showing that it takes constant time per operation.

\begin{proof}
The time analysis follows immediately from the description of the data structure update operations. These updates maintain invariant the properties of the set $B$ and the dictionary of sets $C[i]$ that they partition $S$ properly by their values of $f(x)$, that $B$ consists exactly of those elements of $H$ with $f(x)=|H|$, and that $H$ consists of $B$ together with those elements of $S$ with $f(x)>|H|$.

Thus, $h=|H|$ has the property that there exists a set (namely $H$) with $h$ elements, all of which have function value at least $h$. There can be no larger $h'$ with the same property, because all of the elements with value greater than $h$ belong to $H$ already so there can be no larger set of elements with larger values. Thus, $h$ is the correct $h$-index of $S$ and $f$, and $(H,S\setminus H)$ is a correct $h$-partition.
\end{proof}

Next we prove Theorem~\ref{thm:gradual}, the time analysis of our data structure for maintaining a partition of a graph into low and high degree vertices with a very low number of moves of vertices from one part of the partition to the other.

As an accounting technique for the analysis of the algorithm (not something actually stored within our data structure) we associate a (fractional) number of ``credits'' with each member of $P$, that is zero when that element is added to $P$. Each increment operation adds $1/|H|^2$ credit to each current member of $P$, and each decrement operation on a member of $P$ adds $1/|H|$ credits to that member.

\begin{lemma}
\label{lem:quadratic-increments}
Any sequence of operations during which $|H|$ changes from $h$ to $h'>h$ includes at least $(h'-h)^2$ increment operations.
\end{lemma}

\begin{proof}
There exist at least $h'-h$ members of the set $H$ after the sequence that were not members prior to the sequence. Each of these elements has $f(x)\le h$ prior to the sequence (else it would belong to $H$) and $f(x)\ge h'$ after the sequence, so the number of increments for these elements alone must have been at least $(h'-h)^2$.
\end{proof}

\begin{lemma}
\label{lem:deficit-accounting}
Any element $x$ that is removed from $P$ must have accumulated $\Omega(1)$ credits.
\end{lemma}

\begin{proof}
Let $h$ be the value of $|H|$ at the time $x$ was added to $P$, and $h'$ be the value of $\max(h,|H|)$ at the time it is removed. Then by the previous lemma, $x$ must have accumulated $(h'-h)^2/h^2$ credits from increment operations, and $\Omega((2h-h')/h)$ credits from decrement operations. But for any $h'\ge h$, $(h'-h)^2/h^2+(2h-h')/h=\Omega(1)$.
\end{proof}

The proof of Theorem~\ref{thm:gradual} now follows.

\begin{proof}
The number of additions is equal to the number of removals, plus the number of items that remain in $H$ at the end of the sequence. But by Lemma~\ref{lem:quadratic-increments} we can find a subsequence $I$ of increase operations such that the final value of $|H|$ is $O(\sum_{i\in I} 1/h_i)$. Thus, we need count only the number of times elements are removed from $P$.
By Lemma~\ref{lem:deficit-accounting}, this number of removals is proportional to the total number of credits that have been accumulated by all elements over the course of $\sigma$. But, since each operation assigned at most $1/h_i$ credits, this total is at most $q$.
\end{proof}

\section*{Appendix II: Additional subgraph counting data structures}

If $s_i=s_i(G)$ denote the number of star subgraphs $K_{1,i}$ in $G$, we may maintain $s_i$, for any constant $i$, in constant time per update, as it is a sum of polynomials of the vertex degrees: $s_i=\sum_v d_v (d_v-1)\cdots (d_v-i-1)/i!$. For instance, the number of claws (three-leaf stars) in $G$ is $s_3=\sum_v d_v (d_v-1)(d_v-2)/6$.
In at least one other nontrivial case we may maintain the number of four-vertex subgraphs of a certain type as efficiently as the number of triangles.

\begin{theorem}
We may maintain a dynamic graph subject to edge insertions and deletions and to insertions and deletions of isolated vertices, and keep track of the number $p_3$ of four-vertex paths in the graph, in amortized time $O(h)$ per update where $h$ is the $h$-index of the graph at the time of an update.
\end{theorem}

\begin{proof}
Let $q$ denote the number of sequences of three edges that form either a path or a cycle in $G$. Let $d_v$ denote the degree of $v$ (that is, its number of incident edges), and let $P_v$ denote the number of two-edge paths having $v$ as an endpoint (that is, $\sum(d_w-1)$ where the sum is over all neighbors of $v$ in $G$).

Inserting an edge $uv$ into the graph $G$ increases $q$ by
$d_u d_v+P_u+P_v$: the term $d_u d_v$ counts the paths with $uv$ as middle edge, and the other two terms count the paths having $v$ or $u$ as endpoint. Similarly, removing edge $uv$ decreases $q$ by
$(d_u-1)(d_v-1)+(P_u-d_v+1)+(P_v-d_u+1)$. Thus, if we can calculate $P_u$ and $P_v$, we can correctly update~$q$.

Our data structure stores the numbers $d_v$ for each vertex $v$, and the numbers $P_u$ only for those vertices $u$ that belong to the set $H$ maintained by the gradual partition of Section~\ref{sec:gradual}. When a vertex is added to $H$, the value $P_u$ stored for it may be computed in time $O(h)$. When we insert or delete an edge $uv$, the numbers $P_u$ and $P_v$ that we need to use to update $q$ may be found either by looking them up in this data structure (if the endpoints $u$ or $v$ of the updated edge belong to $H$) or in time $O(h)$ by looking at all neighbors of the endpoints if they do not belong to $H$. Finally, whenever we insert or delete an edge $uv$, we must update the numbers $P_w$ for all vertices $w$ belonging to $H$, where either $w$ is one of the two endpoints $u$ and $v$ or it is adjacent to one or both of these endpoints; this update may be performed in constant time per member of $H$, or $O(h)$ time total.

The number of four-vertex paths that we maintain is then $p_3=q-3c_3$ where $c_3$ denotes the number of triangles in the graph as maintained by our other structures.
\end{proof}

The counts of larger subgraphs in $G$ obey additional linear relations: for instance, $\sum_v P_v^2 = p_4 + 2 p_2 + 3 s_3 + 4 c_4$. However we have not been able to exploit these relations by finding efficient algorithms for maintaining the quantities $p_4$ and $c_4$.

\vfill\eject
\section*{Appendix III: Detailed analysis of real-world network data}

\begin{figure}[hbt]
	\centering
		\includegraphics[scale=.4]{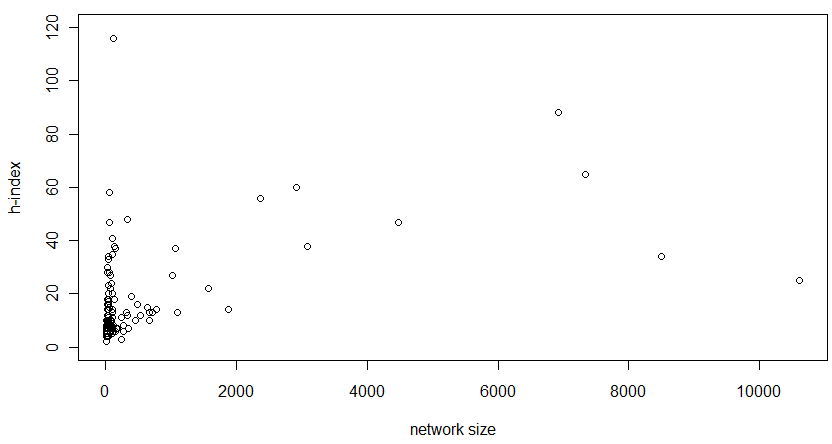}
	\caption{Scatter plot of $h$-index and network size}
	\label{fig:Nvh}
\end{figure}

We calculated the $h$-index of the networks in our sample in R, using a subroutine provided by Carter Butts. The data that results from this calculation in plotted in Figure \ref{fig:Nvh}.

\begin{figure}[p]
	\centering
		\includegraphics[scale=.3]{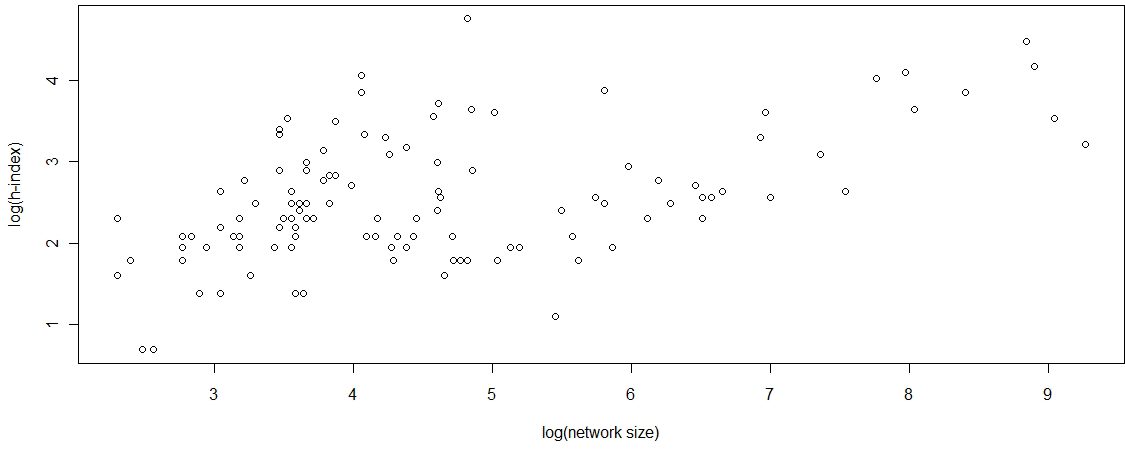}
	\caption{Scatter plot of $h$-index and network size, on log-log scale}
	\label{fig:log(size).log(hindex)}
\end{figure}

Figure \ref{fig:Nvh} suggests that the data might be more appropriately viewed on a log-log scale. This plot is seen in Figure \ref{fig:log(size).log(hindex)}. 

\subsection{Quantile regression}

To find an upper bound on the scaling of the $h$-index of our real world networks we clustered the data into two groups, and used quantile regression to fit the data with curves of the form $\log h=\beta_0+\beta_1\log n$, at the 95th percentile. That is, we are looking for a power law $h=cn^{\beta_1}$, and we want 95\% of the graphs to have an $h$-index no larger than the one predicted by this law. We fit a law of this type to the two clusters separately to provide a more conservative and substantive prediction. The resulting regression lines are reported in Table \ref{tab:regrslt}. Corresponding goodness of fit measure are also reported in Table \ref{tab:gof}. We note that these are conservative estimates and the actual scaling is likely better.

\begin{table}[p]
	\centering
		\begin{tabular}{|c|c|c|c|c|}
		\hline
		 Cluster & Intercept $\beta_0$ & Slope $\beta_1$ & df\\
		\hline
		\hline
1 & 0.0609 &   0.9735 &  92 \\
&\footnotesize (-0.964, 2.581) & \footnotesize(0.231, 1.266)&\\
\hline
2 & -0.598  &  0.604  & 44   \\
& \footnotesize(-1.938, 5.248)& \footnotesize(0.44712, 0.847)& \\
\hline	
		\end{tabular}
         \smallskip
	\caption{Coefficients for quantile regression lines}
	\label{tab:regrslt}
\end{table}

\begin{center}
\begin{figure}[p]
\centering
		\includegraphics[scale=.3]{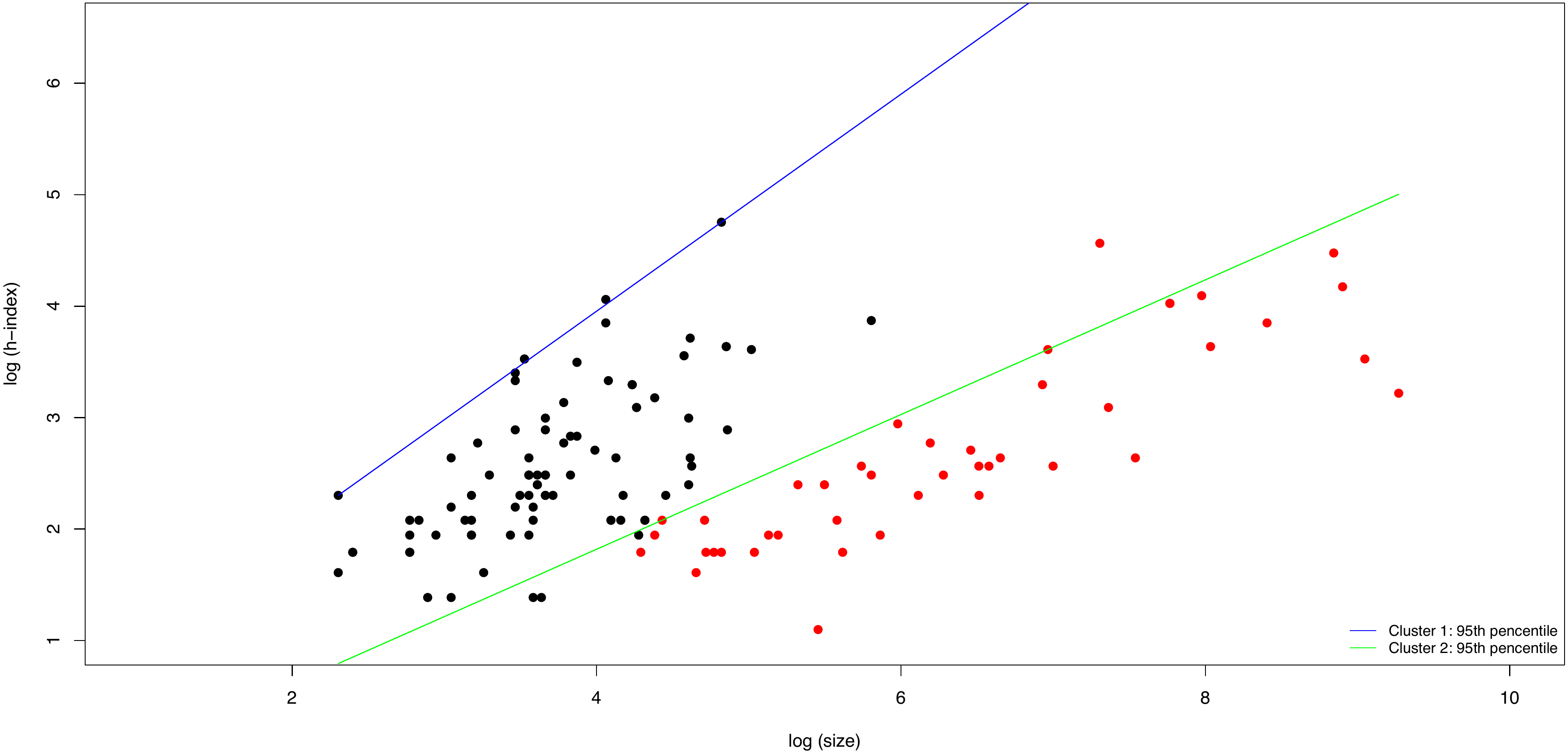}
	\caption{H-index scaling using quantile regression fits}
	\label{quantreg}
\end{figure}
\end{center}

\begin{table}[p]
	\centering
		\begin{tabular}{|c|c|c|c|c|}
		\hline
		 Cluster & log-like & AIC & BIC\\
		\hline
		\hline
1 & -109.345  &   222.691 &  227.734 \\
2 & -41.071  &  86.143  & 89.712   \\
\hline	
		\end{tabular}
         \smallskip
	\caption{Goodness of fit measures for quantile regression lines}
	\label{tab:gof}
\end{table}

\vfill\eject
\section*{Appendix IV: Raw data from analysis of real-world networks}

\centering
\begin{minipage}[b]{0.45\linewidth}\centering
\begin{tabular}{|@{~}r@{~}|@{~}r@{~}|@{~}c@{~}|@{~}c@{~}|@{~}c@{~}|} \hline
$n$ & $h$ & $\log n$ & $\log h$ & $\displaystyle\frac{\log h}{\log n}$ \\
\hline
10&5&2.3026&1.6094&0.6990\\ 
10&10&2.3026&2.3026&1.0000\\ 
11&6&2.3979&1.7918&0.7472\\ 
11&6&2.3979&1.7918&0.7472\\ 
12&2&2.4849&0.6931&0.2789\\ 
13&2&2.5649&0.6931&0.2702\\ 
16&6&2.7726&1.7918&0.6462\\ 
16&6&2.7726&1.7918&0.6462\\ 
16&8&2.7726&2.0794&0.7500\\ 
16&7&2.7726&1.9459&0.7018\\ 

17&8&2.8332&2.0794&0.7340\\ 
18&4&2.8904&1.3863&0.4796\\ 
19&7&2.9444&1.9459&0.6609\\ 
21&14&3.0445&2.6391&0.8668\\ 
21&9&3.0445&2.1972&0.7217\\ 
21&4&3.0445&1.3863&0.4553\\ 
23&8&3.1355&2.0794&0.6632\\ 
24&10&3.1781&2.3026&0.7245\\ 
24&8&3.1781&2.0794&0.6543\\ 
24&7&3.1781&1.9459&0.6123\\ 

24&7&3.1781&1.9459&0.6123\\ 
25&16&3.2189&2.7726&0.8614\\ 
26&5&3.2581&1.6094&0.4940\\ 
27&12&3.2958&2.4849&0.7540\\ 
31&7&3.4340&1.9459&0.5667\\ 
32&9&3.4657&2.1972&0.6340\\ 
32&28&3.4657&3.3322&0.9615\\ 
32&30&3.4657&3.4012&0.9814\\ 
32&18&3.4657&2.8904&0.8340\\ 
33&10&3.4965&2.3026&0.6585\\ 

34&34&3.5264&3.5264&1.0000\\ 
34&34&3.5264&3.5264&1.0000\\ 
35&10&3.5553&2.3026&0.6476\\ 
35&12&3.5553&2.4849&0.6989\\ 
\hline
\end{tabular}
\end{minipage}
\begin{minipage}[b]{0.45\linewidth} \centering
\begin{tabular}{|@{~}r@{~}|@{~}r@{~}|@{~}c@{~}|@{~}c@{~}|@{~}c@{~}|} \hline
$n$ & $h$ & $\log n$ & $\log h$ & $\displaystyle\frac{\log h}{\log n}$ \\
\hline
35&12&3.5553&2.4849&0.6989\\ 
35&7&3.5553&1.9459&0.5473\\ 
35&14&3.5553&2.6391&0.7423\\ 
35&12&3.5553&2.4849&0.6989\\ 
36&4&3.5835&1.3863&0.3869\\ 
36&9&3.5835&2.1972&0.6131\\ 
36&8&3.5835&2.0794&0.5803\\ 
37&11&3.6109&2.3979&0.6641\\ 
37&11&3.6109&2.3979&0.6641\\ 
37&12&3.6109&2.4849&0.6882\\ 

38&4&3.6376&1.3863&0.3811\\ 
39&10&3.6636&2.3026&0.6285\\ 
39&10&3.6636&2.3026&0.6285\\ 
39&12&3.6636&2.4849&0.6783\\ 
39&18&3.6636&2.8904&0.7890\\ 
39&20&3.6636&2.9957&0.8177\\ 
39&12&3.6636&2.4849&0.6783\\
41&10&3.7136&2.3026&0.6200\\ 
44&16&3.7842&2.7726&0.7327\\ 
44&23&3.7842&3.1355&0.8286\\ 

46&12&3.8286&2.4849&0.6490\\ 
46&17&3.8286&2.8332&0.7400\\ 
48&33&3.8712&3.4965&0.9032\\ 
48&33&3.8712&3.4965&0.9032\\ 
48&17&3.8712&2.8332&0.7319\\ 
54&15&3.9890&2.7081&0.6789\\ 
58&47&4.0604&3.8501&0.9482\\ 
58&58&4.0604&4.0604&1.0000\\ 
59&28&4.0775&3.3322&0.8172\\ 
60&8&4.0943&2.0794&0.5079\\ 

60&8&4.0943&2.0794&0.5079\\ 
62&14&4.1271&2.6391&0.6394\\ 
64&8&4.1589&2.0794&0.5000\\ 
65&10&4.1744&2.3026&0.5516\\ 
\hline
\end{tabular}
\end{minipage}

\vfill\eject

\centering
\begin{minipage}[b]{0.45\linewidth}\centering
\begin{tabular}{|@{~}r@{~}|@{~}r@{~}|@{~}c@{~}|@{~}c@{~}|@{~}c@{~}|} \hline
$n$ & $h$ & $\log n$ & $\log h$ & $\displaystyle\frac{\log h}{\log n}$ \\
\hline
69&27&4.2341&3.2958&0.7784\\ 
69&27&4.2341&3.2958&0.7784\\ 
69&27&4.2341&3.2958&0.7784\\ 
71&22&4.2627&3.0910&0.7251\\ 
71&22&4.2627&3.0910&0.7251\\ 
72&7&4.2767&1.9459&0.4550\\ 
73&6&4.2905&1.7918&0.4176\\ 
75&8&4.3175&2.0794&0.4816\\ 
75&8&4.3175&2.0794&0.4816\\ 
80&7&4.3820&1.9459&0.4441\\ 

80&24&4.3820&3.1781&0.7252\\ 
84&8&4.4308&2.0794&0.4693\\ 
86&10&4.4543&2.3026&0.5169\\ 
97&35&4.5747&3.5553&0.7772\\ 
97&35&4.5747&3.5553&0.7772\\ 
100&11&4.6052&2.3979&0.5207\\ 
100&20&4.6052&2.9957&0.6505\\ 
101&14&4.6151&2.6391&0.5718\\ 
101&41&4.6151&3.7136&0.8047\\ 
102&13&4.6250&2.5649&0.5546\\ 

105&5&4.6540&1.6094&0.3458\\ 
111&8&4.7095&2.0794&0.4415\\ 
112&6&4.7185&1.7918&0.3797\\ 
118&6&4.7707&1.7918&0.3756\\ 
124&116&4.8203&4.7536&0.9862\\ 
124&6&4.8203&1.7918&0.3717\\ 
128&38&4.8520&3.6376&0.7497\\ 
128&38&4.8520&3.6376&0.7497\\ 
128&38&4.8520&3.6376&0.7497\\ 
129&18&4.8598&2.8904&0.5947\\

151&37&5.0173&3.6109&0.7197\\ 
154&6&5.0370&1.7918&0.3557\\ 
169&7&5.1299&1.9459&0.3793\\ 
180&7&5.1930&1.9459&0.3747\\ 
\hline
\end{tabular}
\end{minipage}
\begin{minipage}[b]{0.45\linewidth} \centering
\begin{tabular}{|@{~}r@{~}|@{~}r@{~}|@{~}c@{~}|@{~}c@{~}|@{~}c@{~}|} \hline
$n$ & $h$ & $\log n$ & $\log h$ & $\displaystyle\frac{\log h}{\log n}$ \\
\hline
205&11&5.3230&2.3979&0.4505\\ 
234&3&5.4553&1.0986&0.2014\\ 
244&11&5.4972&2.3979&0.4362\\ 
265&8&5.5797&2.0794&0.3727\\ 
275&6&5.6168&1.7918&0.3190\\ 
311&13&5.7398&2.5649&0.4469\\ 
332&48&5.8051&3.8712&0.6669\\ 
332&12&5.8051&2.4849&0.4281\\ 
352&7&5.8636&1.9459&0.3319\\ 
395&19&5.9789&2.9444&0.4925\\ 

452&10&6.1137&2.3026&0.3766\\ 
489&16&6.1924&2.7726&0.4477\\ 
533&12&6.2785&2.4849&0.3958\\ 
638&15&6.4583&2.7081&0.4193\\ 
673&13&6.5117&2.5649&0.3939\\ 
674&10&6.5132&2.3026&0.3535\\ 
719&13&6.5779&2.5649&0.3899\\ 
775&14&6.6529&2.6391&0.3967\\ 
1022&27&6.9295&3.2958&0.4756\\ 
1059&37&6.9651&3.6109&0.5184\\ 

1096&13&6.9994&2.5649&0.3665\\ 
1490&96&7.3065&4.5643&0.6247\\ 
1577&22&7.3633&3.0910&0.4198\\ 
1882&14&7.5401&2.6391&0.3500\\ 
2361&56&7.7668&4.0254&0.5183\\ 
2361&56&7.7668&4.0254&0.5183\\ 
2361&56&7.7668&4.0254&0.5183\\ 
2909&60&7.9756&4.0943&0.5134\\ 
3084&38&8.0340&3.6376&0.4528\\ 
4470&47&8.4051&3.8501&0.4581\\ 

6927&88&8.8432&4.4773&0.5063\\ 
7343&65&8.9015&4.1744&0.4690\\ 
8497&34&9.0475&3.5264&0.3898\\ 
10616&25&9.2701&3.2189&0.3472\\ 
\hline
\end{tabular}
\end{minipage}

\end{document}